\documentclass[review,number,sort&compress]{elsarticle}
\usepackage{graphicx}

\usepackage{amssymb}

\usepackage{amsthm}

\usepackage{makecell}

\usepackage{amsmath}

\usepackage[utf8]{inputenc}

\usepackage{lineno}
\modulolinenumbers[1]
\usepackage[english]{babel}

\usepackage{subfigure}
\usepackage{multirow}
\usepackage[colorlinks=true]{hyperref}
\usepackage{breakurl}

\usepackage[usenames, dvipsnames]{color}

\usepackage{xcolor}

\usepackage{colortbl}

\usepackage{dblfloatfix}

\usepackage{rotating}

\usepackage{microtype}

\journal{Nuclear Instruments and Methods in Physics Research, Section A: Accelerators, Spectrometers, Detectors and Associated Equipment}

\begin{document}

\begin{frontmatter}

\title{Characterization of a CLYC detector for underground experiments}

\author[ciemat]{T. Martinez \corref{mycor1}}
\cortext[mycor1]{Corresponding author.}
\ead{trino.martinez@ciemat.es}
\author[ciemat]{A. P\'{e}rez de Rada}
\author[ciemat]{D. Cano-Ott}
\author[ciemat]{R. Santorelli}
\author[lsc]{I. Bandac}
\author[ciemat]{P. Garcia Abia}
\author[ciemat]{A.R. Garcia}
\author[lsc]{A. Ianni}
\author[ciemat]{B. Montes}
\author[ciemat]{L. Romero}
\author[ciemat]{D. Villamarin}

\address[ciemat]{Centro de Investigaciones Energ\'{e}ticas, Medioambientales y Tecnol\'{o}gicas (CIEMAT), Avenida Complutense 40, Madrid 28040, Spain}
\address[lsc]{Laboratorio Subterr\'{a}neo de Canfranc (LSC), Paseo de los Ayerbes s/n, Canfranc-Estaci\'{o}n 22880, Spain}

\begin{abstract}

Large size detectors based on Cs$_{2}$LiYCl$_{6}$:Ce (CLYC) are capable of performing a combined $\gamma$-ray and neutron spectrometry and constitute a promising technology for a wide range of applications in nuclear and high energy physics. Due to their novelty, the comprehensive characterization of the performance of individual CLYC detectors is of great importance for determining their range of applicability. In this work we report on a wide series measurements performed with a commercial 2"x2" CLYC crystal. Good energy and timing resolution values of 4.7$\%$ (@ 662 keV) and 1340 ps (FWHM), respectively, were achieved, and a neutron/$\gamma$ separation figure of merit value of 4.2 was obtained. A dedicated measurement for investigating the intrinsic background of the detector was performed at the Laboratorio Subterr\'{a}neo de Canfranc (Spain). It evidenced a sizeable contamination in the detector materials which poses limits in the use of CLYC in low background experiments. In addition, detailed Monte Carlo simulations with the GEANT4 toolkit were performed for modeling the response function of the CLYC detector to $\gamma$-rays. An excellent agreement with the experimental data has been achieved.
\end{abstract}


\end{frontmatter}


\section{Introduction}

In the last decade, the interest on the Cs$_{2}$LiYCl$_{6}$:Ce (CLYC) scintillators has increased due to their enormous potential for a broad range of applications that benefit from its combined $\gamma$-ray and neutron radiation~\cite{Glo1,Bud,McD} detection properties. As a neutron detector, CLYC allows the detection of thermal neutrons through the $^{6}$Li(n$_{th},\alpha$)$^{3}$H reaction with efficiencies comparable to $^{3}$He tubes of similar size, and fast neutrons via the $^{35}$Cl(n,p)$^{35}$S and other neutron-induced charged particle production reactions with sufficiently large cross-sections ~\cite{Oly1, Smi}.

As a $\gamma$-ray detector, CLYC features a good energy resolution (typically 5$\%$ FWHM at 662 keV) and a good proportionality between deposited energy and light output in the range from few keV to several MeV, better than other commonly used $\gamma$-ray scintillators such as NaI(Tl) and CsI(Tl). In addition, it offers a very effective particle identification via pulse shape discrimination (PSD) techniques due to the distinctive decay times and magnitudes of the light signal components for charged particle (i.e. secondary particles produced in neutron induced reactions) and $\gamma$-ray interactions.

These properties make CLYC a very attractive candidate for different applications in the field of experimental nuclear, particle and astro-particle physics. A collaboration between the Laboratorio Subterr\'{a}neo de Canfranc (Canfranc underground laboratory - LSC) and CIEMAT (Centro de Investigaciones Energ\'{e}ticas, Medioambientales y Tecnol\'{o}gicas) has been established for investigating the suitability of CLYC in nuclear physics and underground experiments.

In this paper we present a thorough characterization of a CLYC detector produced by Radiation Monitoring Devices Inc. (RMD)~\cite{RMD}. The main characteristics such as the linearity, energy and timing resolution have been determined with both analogue and fully digital electronic chains. We have performed as well extensive simulations with the GEANT4 simulation toolkit~\cite{Ago} of the response of the CLYC to $\gamma$-rays emitted by calibration sources with known activities. The intrinsic efficiency, total and peak have been determined from simulations and compared to experimental values. The quality of the neutron/$\gamma$ separation has also been investigated with a $^{252}$Cf source. Last, the intrinsic background of the CLYC has been measured at the LSC facility.

\section{Experimental details}

\subsection{The detector assembly}

The performance of a commercial 2"$\times$2" CLYC-50-PHI-50-P-118 detector from RMD purchased by the LSC has been investigated at CIEMAT. The crystal was doped with less than $1\%$ Ce$^{3+}$ and enriched to $95\%$ in $^{6}$Li to maximize the sensitivity to thermal neutrons. The detector consists of a 48 mm $\Phi$ $\times$ 50 mm thick cylindrical crystal encapsulated in two aluminum housings. The crystal is wrapped with Teflon and hermetically sealed inside an aluminum casing with a quartz window. The casing is coupled to a Hamamatsu R6233-100 super bialkali (SBA) photomultiplier tube, and the whole detector is packaged into a second aluminum housing. 

The PMT is wired for positive bias voltage and the PMT base is a D-type socket assembly E1198-27 from Hamamatsu. The operational voltage of the PMT is limited up to 1500V, and due to the relatively low good photon yield of the CLYC (20000 ph/MeV) the working voltage has been set below +1450V in order to have good signal-to-noise ratio with reasonably large signal amplitudes while keeping the PMT in a safe conditions.   

\subsection{Description of the measurements}

In order to characterize the detector response, a first set of measurements was carried out with standard $\gamma$-ray calibration sources. We have determined the linearity, energy and time resolutions, as well as the dependence of the gain on the irradiation point, temperature and counting rate in the detector. The response function to $\gamma$-rays and its intrinsic efficiency has also been determined up to 2 MeV. A second set of measurements with a 10 kBq $^{252}$Cf neutron source was performed for studying the quality of the neutron/$\gamma$ separation. Last, but not least, the intrinsic background of the detector has been evaluated through a set of background measurements performed on surface and at the LSC at an overburden of about 2500 m.w.e. (meters of water equivalent).

\begin{figure}
\centering
\includegraphics[width=0.5\columnwidth]{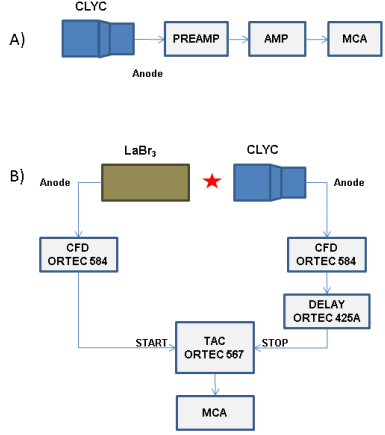}
\caption{Schemes of the electronic setups employed for A) collecting pulse height spectra and B) for measuring the time resolution.}
\label{elec}
\end{figure}

For these measurements, the PMT anode signals have been processed with both analogue and digital electronics. A schematic picture of the analogue electronic chain is shown in Fig.~\ref{elec}A, which has been used for pulse height spectra measurement. The anode signal was sent to a Canberra 2005 preamplifier and the output signal was fed into a Canberra 2020 spectroscopy amplifier. The best results for the energy resolution were obtained with a 6 $\mu$s shaping time. The pulse height spectra of the amplified signals were recorded with an Atomki Palmtop multi-channel analyzer~\cite{mca}.

The time resolution was measured with the time coincidence method. The setup consisted of a LaBr$_{3}$ reference detector, with a fast time response, and the CLYC detector, placed at 180$^{\circ}$ (see Fig.~\ref{elec}B). The $\gamma$-rays emitted within 1 ps from the $^{60}$Co source were detected in both detectors. The anode signals from both PMT (see Fig.~\ref{elec}B)were plugged into an ORTEC 584 constant fraction discriminators and their outputs to an ORTEC 567 Time-to-Amplitude Converter (TAC). The electronic branch of the CLYC included an ORTEC 425A module, adding a known delay between start and stop signals. Finally, the TAC output signal, whose amplitude is proportional to the time difference between detector input pulses, was sent to the MCA. 

Digital electronics have also been used to characterize the neutron/$\gamma$ pulse shape discrimination performance. The anode signals were directly fed into a 4-channel fast digitizer ADQ14DC from SP devices~\cite{SP}. The pulses were recorded at a sampling rate of 1GS/s and 14 bits vertical resolution. The acquisition time window was 16 $\mu$s for each pulse and a trigger threshold was set at -10 mV enough to avoid noise triggering. 

\subsection{Digital pulse shape analysis}

The pulse shape analysis was performed with a custom-made program written in C++. The baseline calculation, pulse detection, amplitude, time and pulse integration in different time windows was performed with dedicated algorithms. In particular, two different integration algorithms have been used to reconstruct the energy spectrum: i) the charge integration of the pulse in an 8 $\mu$s windows starting at the beginning of the pulse, and ii) a shaping algorithm based on a CR-(RC)$^{4}$ filter network~\cite{Nak} with a 6 $\mu$s time constant. The output response of the later is almost Gaussian in shape, having the advantage of a better noise-to-signal performance compared with the charge integration method. The data analysis of the physical parameters such as the amplitude, time and charge, among others, was performed with dedicated software based on the ROOT package~\cite{ROOT}.

\section{Monte Carlo simulation with GEANT4}

The response function of the CLYC detector to $\gamma$-rays has ben calculated by Monte Carlo simulations with GEANT4 (release 10.03.p01). The GEANT4 toolkit was preferred over other simulation codes used in previous works~\cite{Mac,Kim} because of its built-in nuclear decay models (G4RadioactiveDecay) based on evaluated nuclear data libraries (ENSDF~\cite{Bhat}) and its versatility for simulating complex decay schemes with correlated particle emission. This was of advantage in a first set of simulations aimed at reproducing the experimental measurements with standard $\gamma$-ray sources. Simulations of individual mono-energetic $\gamma$-rays were also performed for calculating the energy dependence of the total and full absorption $\gamma$-ray detection efficiency curves.

\begin{figure}[h!]
\centering
\includegraphics[width=0.5\textwidth]{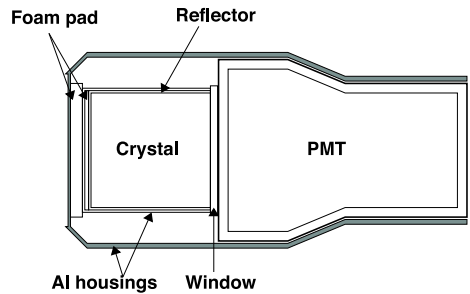}
\caption{Geometry of the CLYC detector implemented in the Monte Carlo simulations.}
\label{g4geom}
\end{figure}

The geometry and material composition of the detector and the rest of the experimental setup (source, detector and source supports and surrounding materials) were implemented to the best of our knowledge. Fig.~\ref{g4geom} illustrates the GEANT4 geometric model of the detector, which corresponds to a faithful reproduction of the drawings provided by the manufacturer. The primary events were generated from an isotropic point source placed in front of the detector, 15 cm away along the symmetry axis. The generation of mono-energetic $\gamma$-rays was isotropic in the simulations for the calculation of the efficiency curves. The well validated Standard Electromagnetic physics package included in GEANT4 was used for defining the electromagnetic interactions. The primary and secondary particles were tracked until they decayed, stopped or escaped the experimental setup. At the end of each simulated event, the energy deposited in the active volume of the detector was integrated. A linear relation between deposited energy and scintillation light was assumed.

\section{Results}
\subsection{Linearity}
\begin{figure}[h!]
\centering
\includegraphics[scale=0.6]{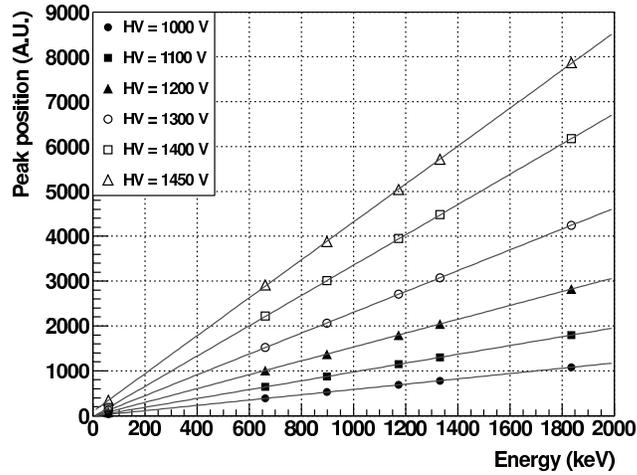}
\caption{Linearity plot for different bias voltages applied to the photomultiplier. The linear fits together with experimental points are shown.}
\label{linearity}
\end{figure}

The linearity of the detector has been investigated in the voltage range between 1000 V and 1450 V. For this purpose, pulse height spectra with $^{60}$Co, $^{88}$Y, $^{137}$Cs and $^{241}$Am sources were recorded with the analogue electronics chain during 600 s collection time per spectrum. The centroid of the $\gamma$-ray peaks versus energy were represented as a function of the bias voltage and a linear function was fitted to each set of data. The different data sets and the results of the fits are shown in Fig.~\ref{linearity}. A good linearity with relative deviations between the points and the fit below 0.6$\%$ has been observed in the entire energy range considered for all the bias voltages.

\subsection{Energy resolution}
\label{eneres}
\begin{figure}[h!]
\centering
\includegraphics[width=0.9\columnwidth]{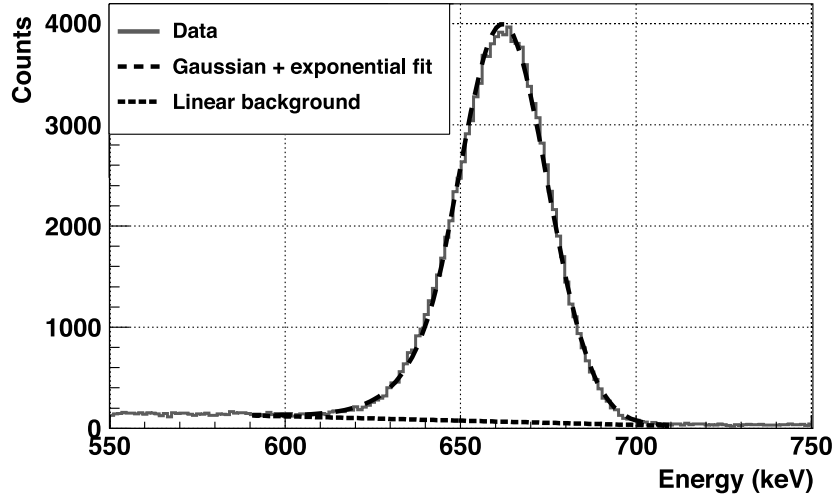}
\caption{Full energy peak for $^{137}$Cs. In solid grey a Gaussian with exponential tail function has been fitted.}
\label{eres_cs}
\end{figure}

It was observed that the $\gamma$-ray peaks in the pulse height spectra are not well described by a Gaussian shape and present a tail at lower energy. For this reason, the spectra were fitted then with a Gaussian + exponential tail function and a linear background. The full width at half maximum (FWHM) was computed from the fit after the background subtraction. Fig.~\ref{eres_cs} shows the pulse height spectrum corresponding to the 662 keV $\gamma$-ray emitted in the $^{137}$Cs decay and the fitted function. As can be seen, the Gaussian + exponential model does not reproduce experimental shape but represents a reasonable model for our purpose. The energy resolution computed in this way amounts to 4.7$\pm1\%$ FWHM. The systematic uncertainty in the value has bee computed from a pure Gaussian fit, which leads to a resolution value of 4.6$\%$, and the resolution computed from the RMS of the data, which amounts to 4.8$\%$. These results are compatible with the values published in literature~\cite{Glo1,Glo13} for crystals of similar size.

\begin{figure}[h!]
\centering
\includegraphics[width=1.\columnwidth]{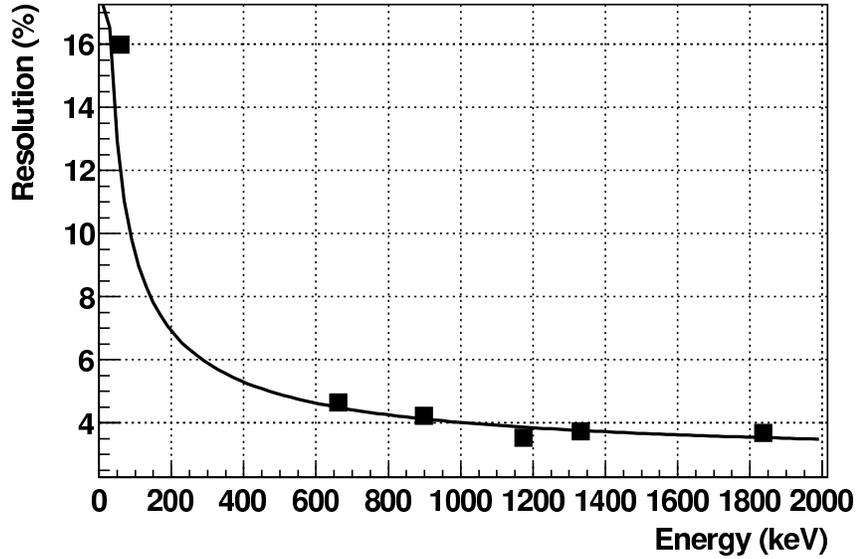}
\caption{Energy resolution as a function of the energy. The squares correspond to the experimental values and the solid line corresponds to a $\Delta E/E$=$\sqrt{a+b/E}$ function fitted to the data.}
\label{energyres}
\end{figure}

The energy resolution was studied as a function of the energy. A two-parameter function $\Delta E/E$=$\sqrt{a+b/E}$ was fitted to the resolution values obtained at various $\gamma$-ray energies. The following fit parameters were obtained: a=7953$\pm$4 and b=8.18$\pm$0.01. Fig.~\ref{energyres} shows the data and the results of the fit. 

\begin{figure}[h!]
\centering
\includegraphics[width=1.\columnwidth]{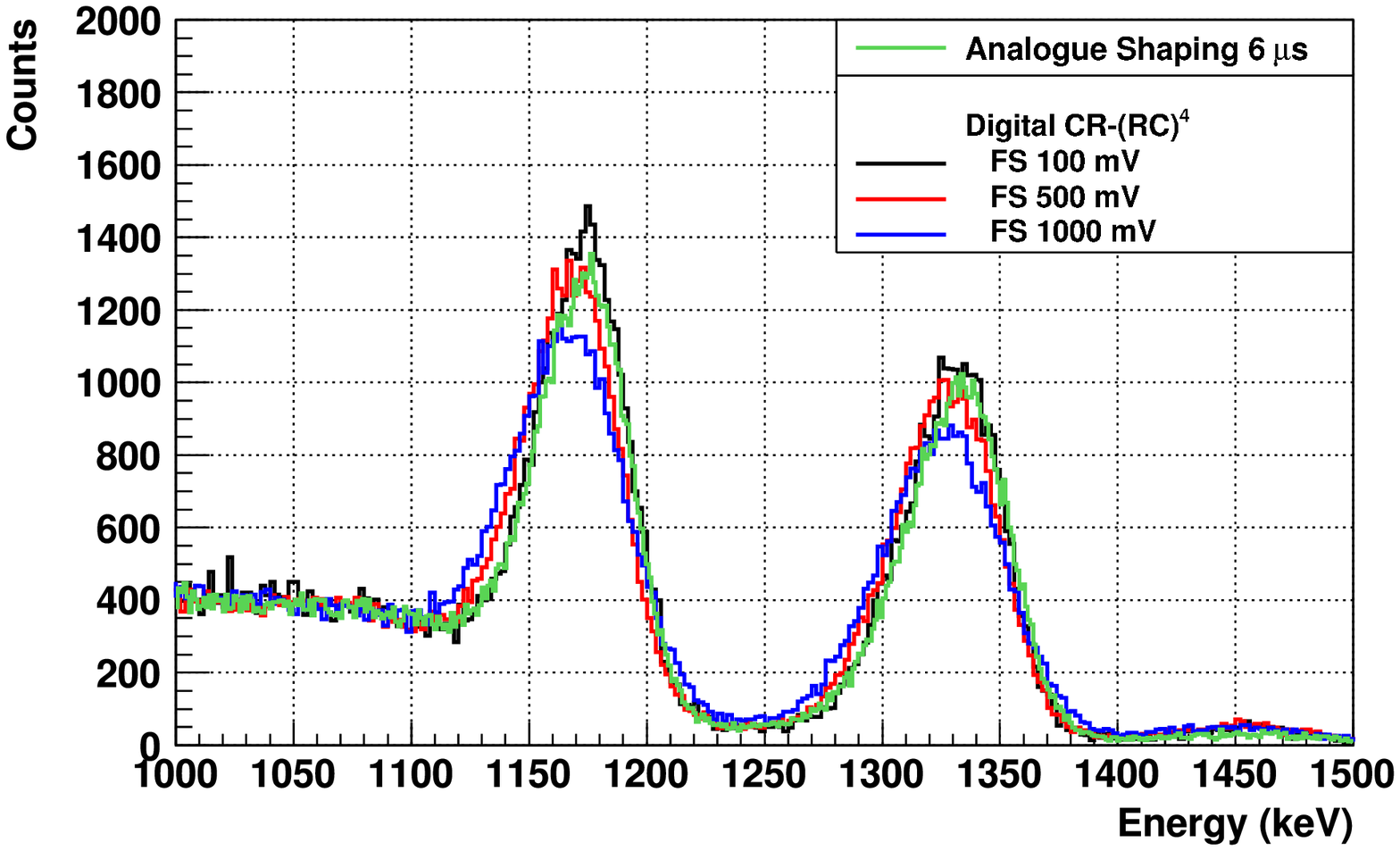} 
\caption{$^{60}$Co energy spectra measured with analogue electronics (green) and with different setting of the vertical scale on the digitizer (black for 100 mV, red for 500 mV and blue for 1000 mV). The CR-(RC)$^{4}$ filter was used to calculate the energy spectra. }
\label{res_rc4}
\end{figure}

The impact of using a fully digital electronic chain and a pulse shape analysis software in the reconstructed energy resolution was also investigated. The PMT signals were recorded with different input scales and sampling rates in the digitizer, in order to quantify the effect of using different values of effective bits and signal points. The corresponding energy spectra were reconstructed as described in section 2.3. Table~\ref{tabres} summarizes the results obtained with each pulse shape analysis algorithm. Better values were obtained when the pulses were digitized with larger number of bits, i.e. smaller full scales. The CR-(RC)$^4$ filter algorithm performed clearly better than the less sophisticated charge integration method and was comparable or even better than analogue electronics, as it is shown in Fig.~\ref{res_rc4}. 
In addition, different sampling rates were applied with the CR-(RC)$^4$ filter with a slight improvement for higher sampling rates.
  
\begin{small}
\begin{table}[h!]
\centering
\begin{tabular}{@{} c @{\kern.7em} c @{\kern1em} c @{\kern0.7em} c @{\kern0.7em} c @{\kern0.7em} c @{}} \hline 
\multicolumn{6}{c}{Direct pulse integration over 8 $\mu$s and 1 GS/s} \\ \hline
\makecell{E$_{\gamma}$ \\(keV)} & Analogue & \makecell{FS 100 \\mV} & \makecell{FS 300\\mV} & \makecell{FS 500 \\mV} & \makecell{FS1000 \\mV} \\ \hline
662  & 4.7(2)$\%$ & 5.4(2)$\%$ & 5.4(2)$\%$ & 5.8(2)$\%$ & 6.8(2)$\%$ \\
1332 & 3.7(2)$\%$ & 4.2(2)$\%$ & 4.2(2)$\%$ & 4.4(2)$\%$ & 4.6(2)$\%$ \\ \hline

\\
\multicolumn{6}{c}{CR-(RC)$^{4}$ filter  ($\tau$=6 $\mu$s) and 1 GS/s} \\ \hline
\makecell{E$_{\gamma}$ \\(keV)} & \makecell{Analogue \\} & \makecell{FS 100 \\mV} & \makecell{FS 300\\mV} & \makecell{FS 500 \\mV} & \makecell{FS1000 \\mV} \\ \hline
662  & 4.7(2)$\%$ & 4.5(2)$\%$ &    & 5.1(2)$\%$ & 6.9(2)$\%$ \\
1332 & 3.7(2)$\%$ & 3.2(2)$\%$ &    & 3.5(2)$\%$ & 4.3(2)$\%$ \\ \hline
\\
\multicolumn{6}{c}{CR-(RC)$^{4}$ filter ($\tau$=6 $\mu$s) and 250 MS/s } \\ \hline
\makecell{E$_{\gamma}$ \\(keV)} & \makecell{Analogue \\} & \makecell{FS 100 \\mV} & \makecell{FS 300\\mV} & \makecell{FS 500 \\mV} & \makecell{FS1000 \\mV} \\ \hline
662  & 4.7(2)$\%$ & 4.5(2)$\%$ &    & 5.2(2)$\%$ & 7.4(2)$\%$ \\
1332 & 3.7(2)$\%$ & 3.2(2)$\%$ &    & 3.5(2)$\%$ & 4.4(2)$\%$ \\ \hline

\end{tabular}
\label{tabres} 
\caption{Energy resolution for different settings of the waveform digitizer. The amplitude of the signals corresponding to 662 and 1332 keV were of 19 and 38 mV respectively }
\end{table}

\end{small}

\subsection{Time resolution}
 
The measurement of the time resolution was first performed with two identical 2"$\times$2"$\times$4" LaBr$_{3}$ crystals coupled to a SBA R6231-100 Hamamatsu PMT with 5 ns. The coincident signals corresponding to the full absorption of the 1173 and 1332 keV $\gamma$-rays (from a $^{60}$Co source) in each detector were processed by the TAC.The calibration of the TAC was performed by adding known delays to the stop signals and determining the change in the centroid of the TAC spectrum. A Gaussian and identical time response for each detector was assumed and the time resolution of an individual LaBr$_{3}$ detector was 354$\pm$4 ps$_{statistical}\pm$10 ps$_{systematic}$, where the statistical uncertainty comes from the fit and the systematic uncertainty comes from the calibration of the TAC.    

\begin{figure}[h!]
\centering
\includegraphics[width=0.9\columnwidth]{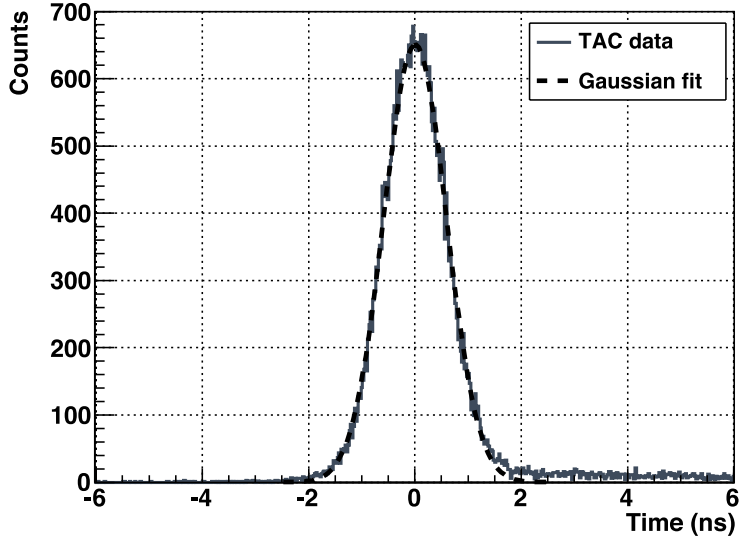}
\caption{Time differences (TAC data) between coincident signals in the CLYC and a LaBr$_{3}$ reference detector. A FWHM resolution of 1385$\pm$6 ps includes the contributions of the individual detectors.}
\label{time}
\end{figure}

The time resolution of the CLYC was measured with respect to the one of the LaBr$_{3}$ detectors, by applying the same method described above. The histogram of the time differences between the coincident CLYC and LaBr$_{3}$ events are shown in Fig.~\ref{time}. The intrinsic time resolution of the CLYC detector has been unfolded from the total response assuming Gaussian responses and a value of 1339$\pm$26 ps, including both statistical (6 ps) and systematic (20 ps) uncertainties.

An equivalent measurement has been performed with digital electronics. The digitized PMT signals were processed with a digital constant fraction (dCFD) algorithm. An intrinsic time resolution of 1510$\pm$30 ps was obtained for the CLYC. We have attributed the slightly worse digital time resolution value to a larger sensitivity of the dCFD to the electronic noise, and don't exclude that better values can be achieved with the addition of some noise filtering procedure. However, the value is comparable to the one obtained by N. D'Olympia, et al.~\cite{Oly2}, who reported an intrinsic time resolution of 1390$\pm$34 ps for a crystal of similar size coupled to a faster PMT with analogue electronics. For this reason, we consider that the use of the SBA R6233-100 PMT with either analogue or digital electronic chains is well suited for a wide range of applications not requiring a sub-nanosecond timing.
 

\subsection{Gain variation}

The gain of scintillation detectors can be noticeably affected by the temperature, in-homogenous light collection or light output and the counting rate. Gain variations induce time dependent shifts in the pulse height spectra which lead to an overall degradation of the energy resolution. If not mitigated, this effect can be especially severe when data need to be collected over long periods, as it is the case in low rate underground experiments. 

\subsubsection*{Gain versus interaction point}

\begin{figure}[h!]
\centering
\subfigure[]{\includegraphics[angle=-90,width=0.50\textwidth]{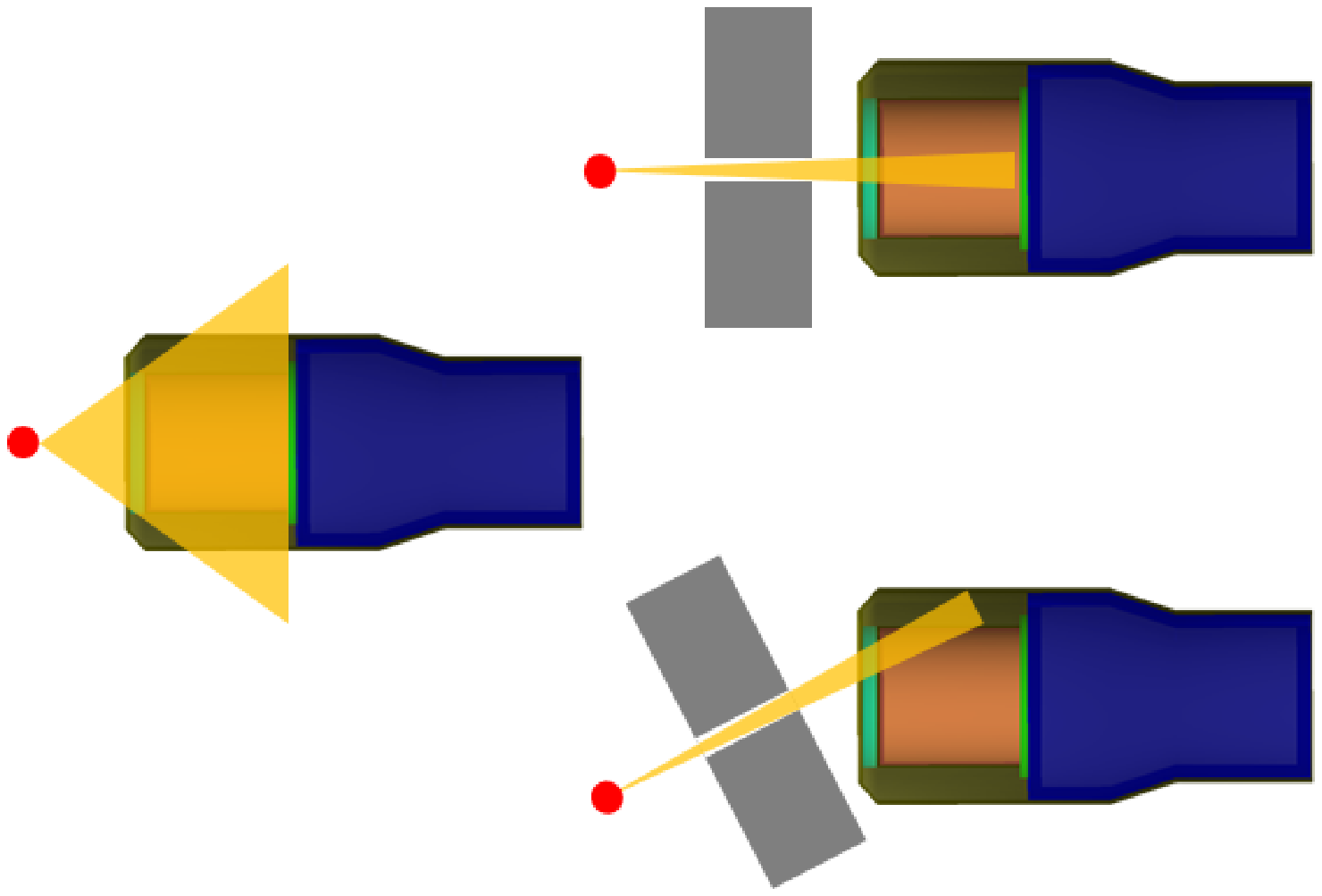}}\label{frontal_irradiation}
\subfigure[]{\includegraphics[angle=-90,width=0.50\textwidth]{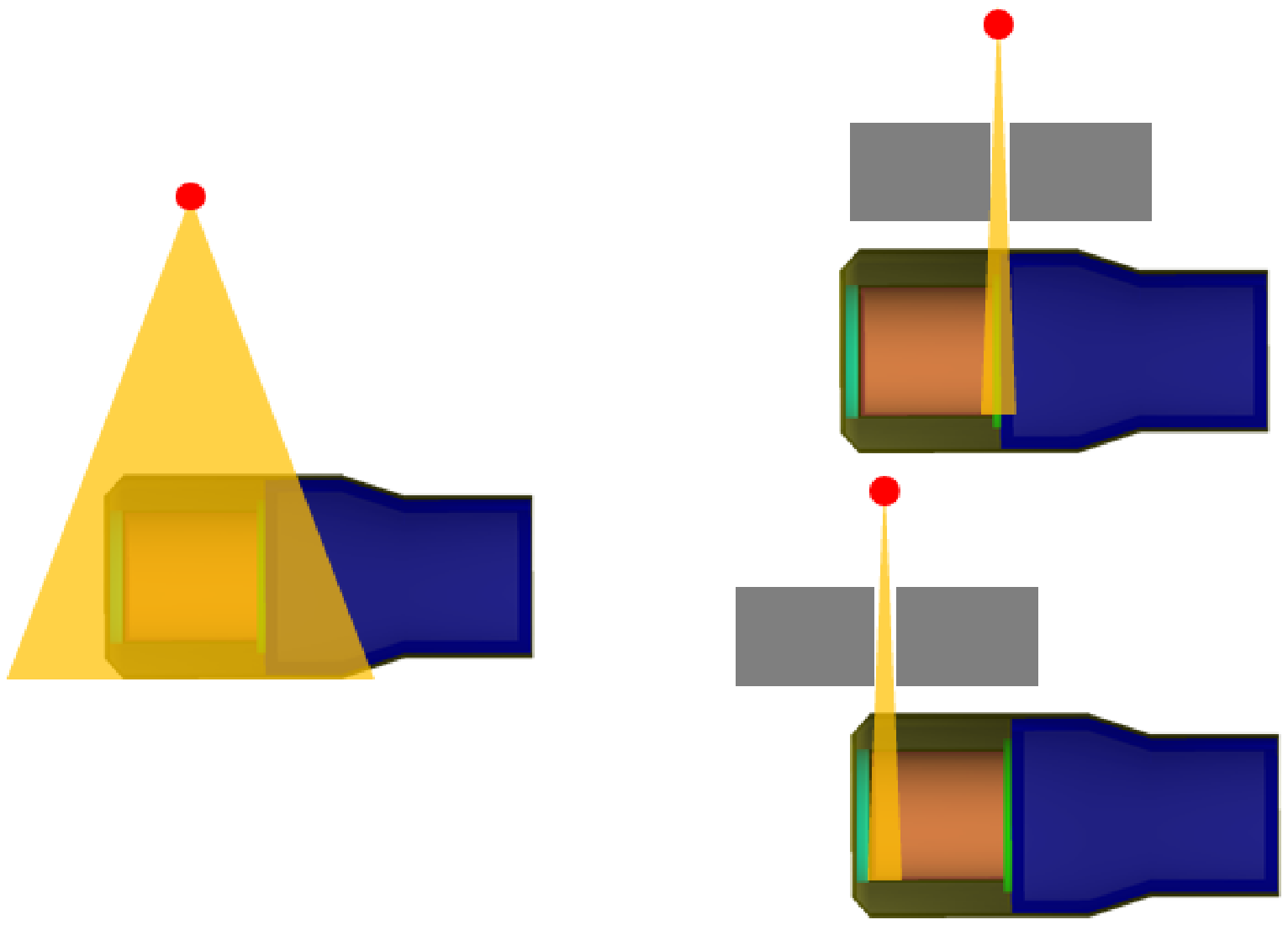}}\label{lateral_irradiation}
\caption{Different experimental configurations for investigating the dependence of the CLYC signal amplitude on the interaction point: a) frontal irradiation with a collimated source pointing at the center of the crystal, pointing to the border of the crystal and with the bare source; b) lateral irradiation with a collimated source close to the PMT window, close to the front side of the crystal and with the bare source.}
\label{irradiation}
\end{figure}

We have investigated the effect of the interaction point on the signal output of the CLYC by irradiating the crystal with a collimated/uncollimated $^{88}$Y source placed at several positions on a) the front face and on the b) lateral surface of the housing. A schematic view of the experimental configurations is depicted in Fig.~\ref{irradiation}.

\begin{figure}[h!]
\centering
\subfigure[Frontal irradiation.]{\includegraphics[width=1.\columnwidth]{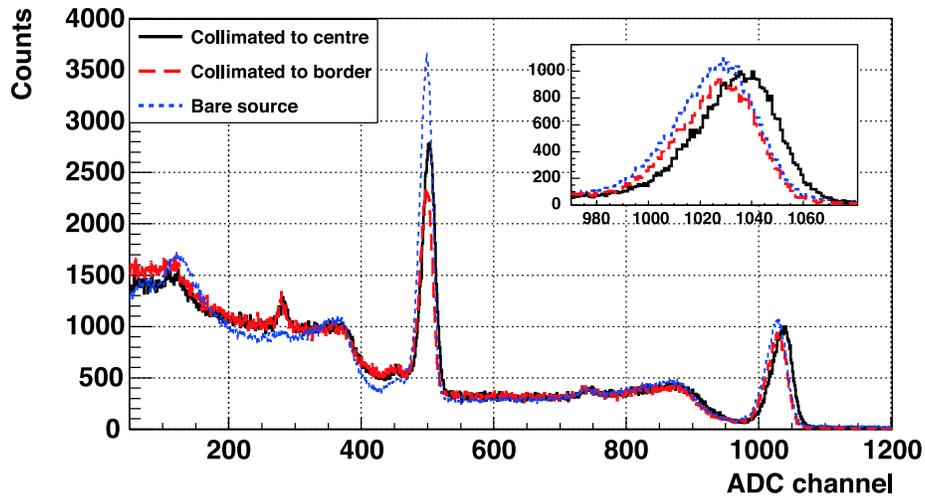}}\label{frontal}
\subfigure[Lateral irradiation.]{\includegraphics[width=1.\columnwidth]{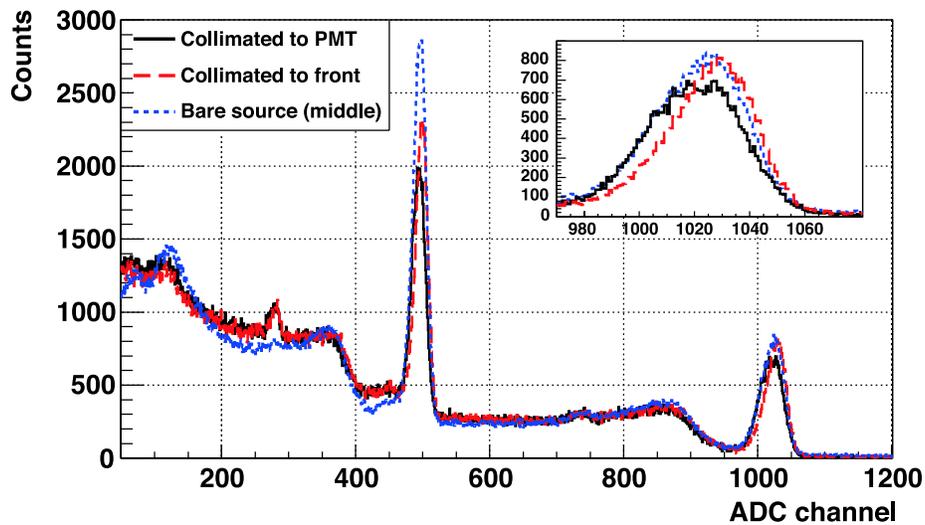}}\label{lateral}
\caption{$^{88}$Y pulse height spectra measured for various irradiation conditions of the detector. The top panel shows the spectra obtained with a frontal irradiation and the bottom panel with a lateral irradiation. The inset shows a zoom on the second peak region.}
\label{lightcoll}
\end{figure}

The spectra recorded from the frontal irradiation are shown in Fig.~\ref{lightcoll} (top panel). The spectra have been normalized to the area of the second peak (1836 keV $\gamma$-ray) for the sake of comparison. Differences in the shape and position of the peaks are observed in the three spectra. This can be seen more clearly in the zoom into the second peak shown in the inset of the figure. In the case of central collimation, the $\gamma$-ray interaction points are concentrated in the region around the detector symmetry axis and the peak is shifted around $~1\%$ toward higher values compared with a frontal bare irradiation where the interaction points will be distributed all around the crystal. With the source collimated to the detector perimeter, the position of the peak is almost the same as in the frontal irradiation. 
A similar effect is observed when the detector is irradiated from the lateral surface: the peak position depends on the distance from the interaction point to the PMT window (Fig.~\ref{lightcoll} bottom panel). In this case lower gain is measured for a collimated irradiation close to the PMT.

The dependence of the gain on the interaction point has been attributed tentatively to an in-homogeneous light collection and/or light yield. However, a more detailed characterization and the simulation of light collection effects in the crystal seem to be necessary for reaching a final conclusion.

\subsubsection*{Gain versus temperature}
\begin{figure}[h!]
\centering
\includegraphics[width=1.\columnwidth]{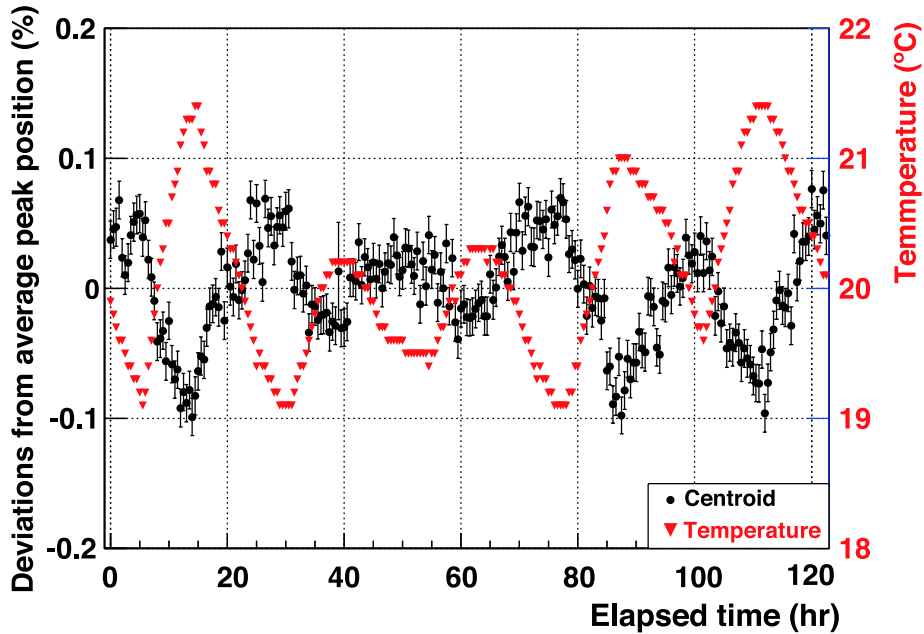}
\caption{Plot of the variation of both the 662 keV relative peak position and the environmental temperature during a five days long measurement. }
\label{temp}
\end{figure}

The effect of the temperature variation on the gain was investigated by irradiating the CLYC detector with a $^{137}$Cs source during a five days long measurement. The temperature in the vincinity of the detector was monitored with a remote sensor. The room was kept under the action of a standard cooling system and the temperature was varying $\pm$1 $^{\circ}$C around 20 $^{\circ}$C. Pulse height spectra with the $^{137}$Cs source were recorded in intervals of five minutes, every 30 minutes along the whole data taking period. The variation in the centroid of the 662 keV peak was used to monitor the detector gain. The results are shown in Fig.~\ref{temp}. The day-night oscillations in the room temperature with a period of 24 hours can seen. It is observed that the relative position of the peak varies inversely with the temperature. As explained in~\cite{Mos}, such behavior can attributed to the negative temperature coefficient of PMT gain and the quantum efficiency (QE) of the photocathode. Relative variation in the peak position amounted to a $\pm$0.1$\%$ for a fluctuation in the room temperature of $\pm$1 $^{\circ}$C. 
For larger variations in the temperature of $\pm$2 $^{\circ}$C, as it was observed when the cooling system was switch-off, the relative variation of the gain increased linearly up to 0.2 $\%$. 

\subsubsection*{Gain versus counting rate}
The variation of the gain with the counting rate has been investigated with a 10 $\mu$Ci $^{207}$Bi source placed a different distances from the detector. The closer to the detector the higher the incident number of $\gamma$-rays and thus the higher counting rate.

\begin{figure}[h!]
\centering
\includegraphics[width=0.9\columnwidth]{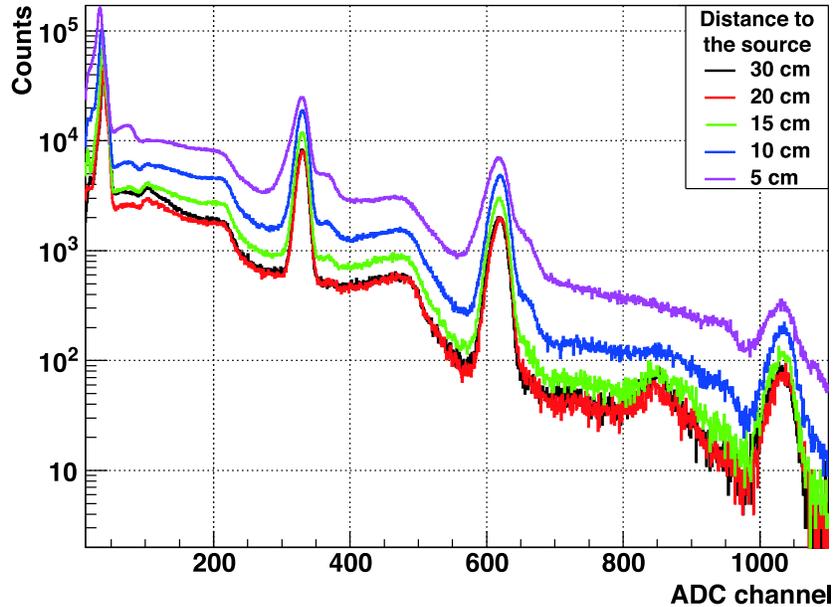}
\caption{Pulse height spectra recorded with the $^{207}$Bi source at different distances from the detector.}
\label{spc_crate}
\end{figure}

\begin{figure}[h!]
\centering
\includegraphics[width=0.9\columnwidth]{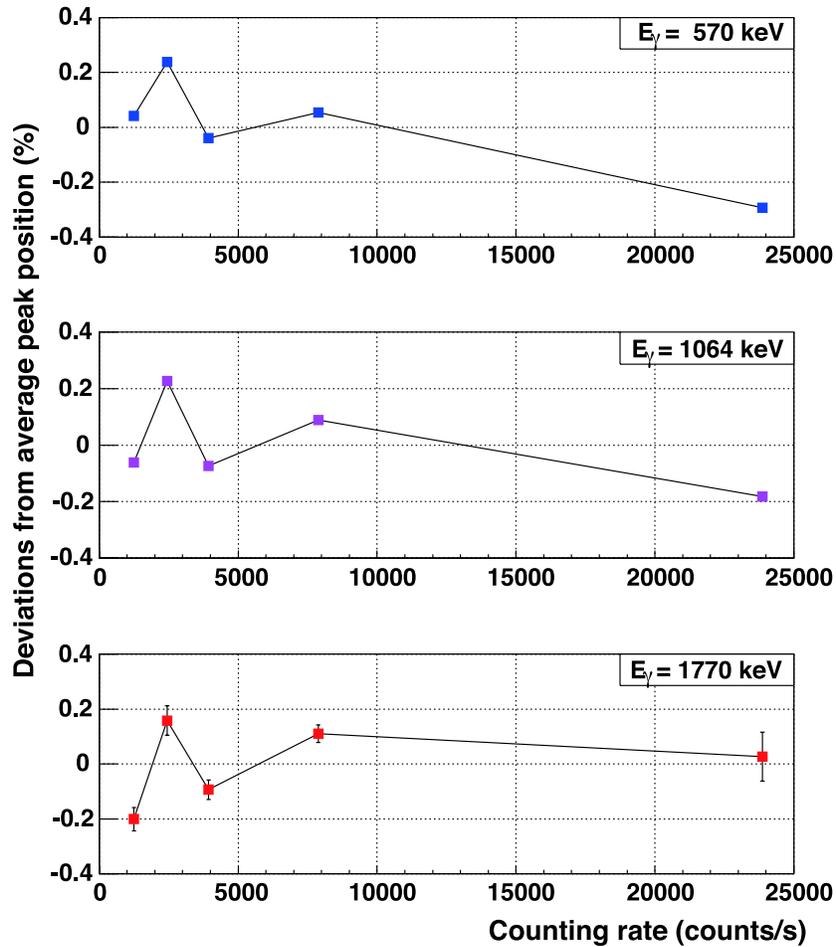}
\caption{Deviations with respect to the average of the $\gamma$-ray peaks from the $^{207}$Bi decay as a function of the counting rate. }
\label{crate}
\end{figure}

The pulse height spectra recorded at different distances are shown in Fig.~\ref{spc_crate}. The average counting rates have been calculated from the integral of the spectrum, after applying the proper dead time correction. The peaks produced by the 570, 1064 and 1770 keV $\gamma$-rays emitted in the decay of $^{207}$Bi and the K X-rays from $^{207}$Pb can be distinguished in the spectrum. Also visible is the effect of the summing with the X-rays, which produces a sum peak overlapping with single $\gamma$-ray peak. The dependence of the peak centroids of the three $\gamma$-ray peaks on the counting rate is plotted in Fig.~\ref{crate}. The conclusion of the analysis is that the variation of the gain is small and below 0.2$\%$ at counting rates as high 25000 counts/s. Each point has systematic uncertainty of 0.1$\%$ due to the distortion induced by the summing, which could not be removed completely with a double peak fitting procedure.
  
\subsection{Response function and efficiency}

\begin{figure}[h!]
\centering
\includegraphics[width=1.\columnwidth]{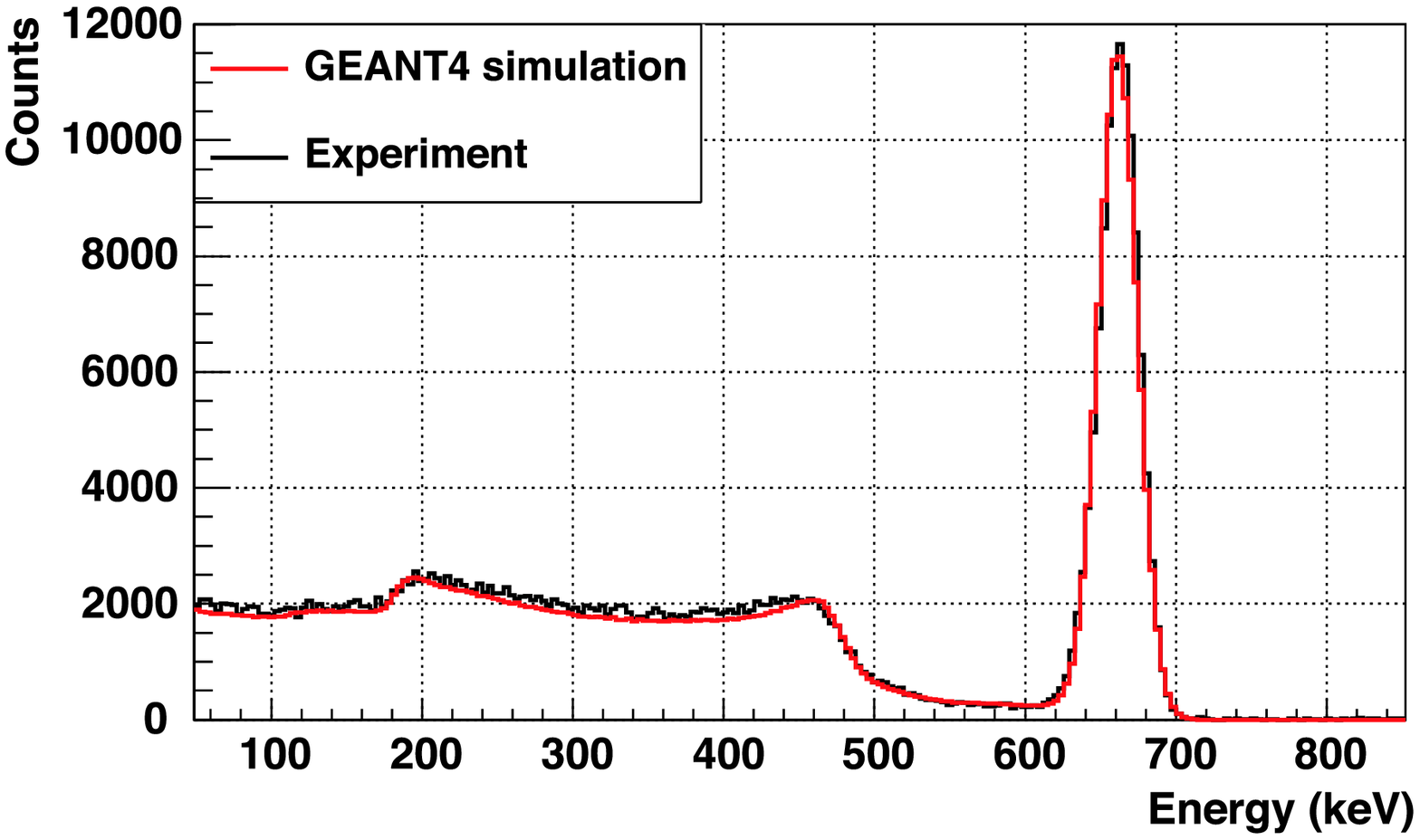}
\caption{Experimental and Monte Carlo simulated $\gamma$-ray response functions for the $^{137}$Cs decay.}
\label{CsMCres}
\end{figure}

\begin{figure}[h!]
\centering
\includegraphics[width=1.\columnwidth]{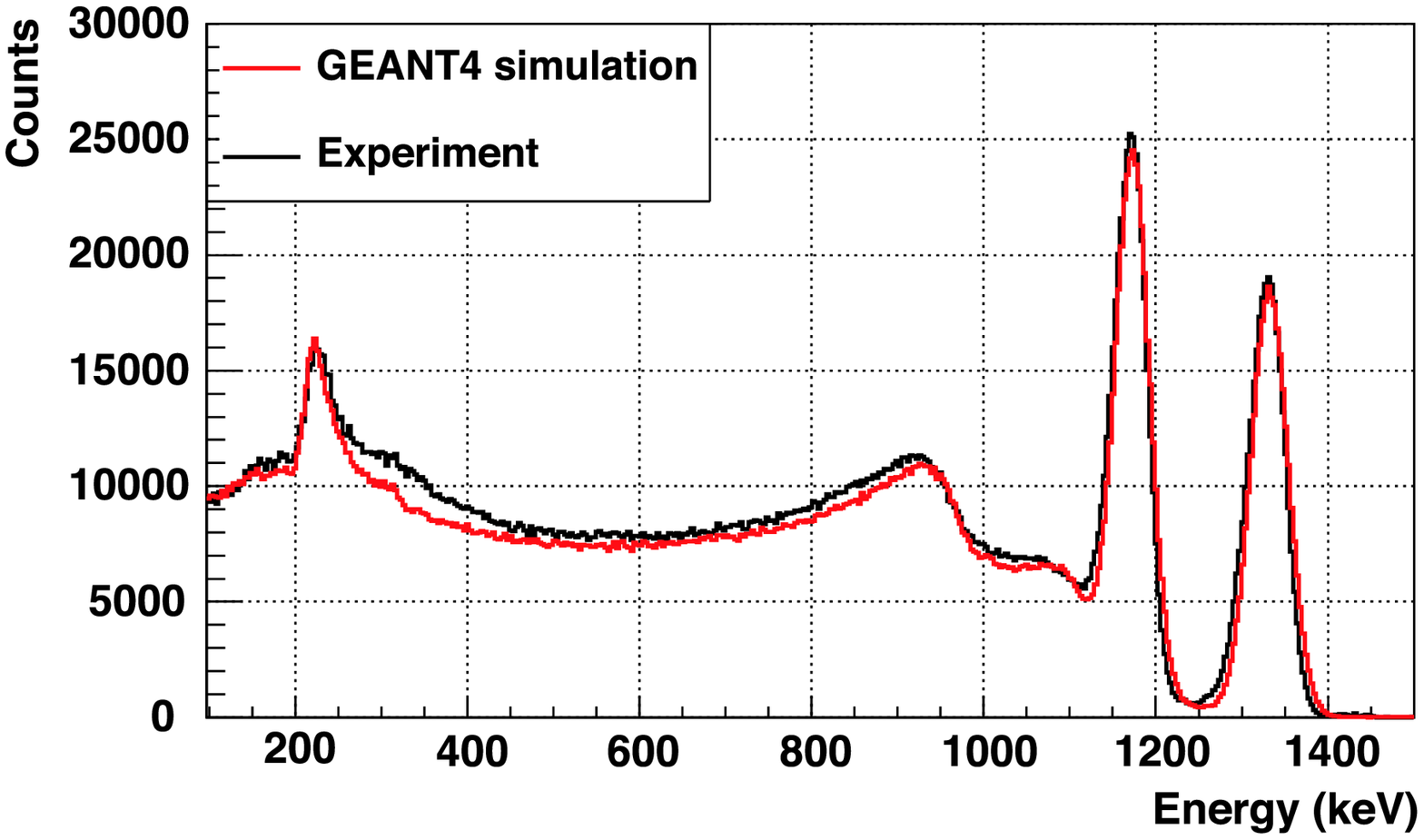}
\caption{Experimental and Monte Carlo simulated $\gamma$-ray response functions for the $^{60}$Co decay.}
\label{CoMCres}
\end{figure}

\begin{figure}[h!]
\centering
\includegraphics[width=1.\columnwidth]{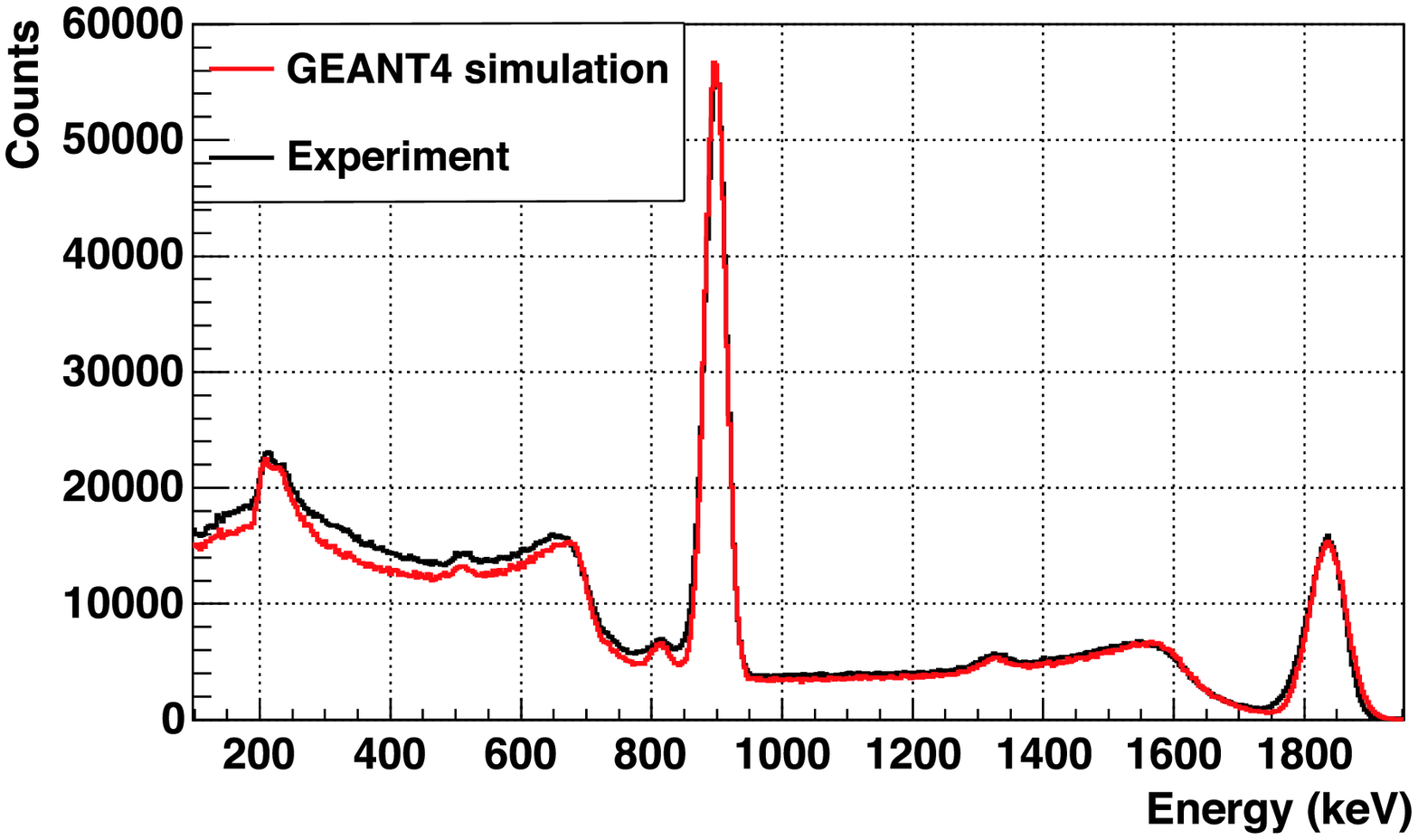}
\caption{Experimental and Monte Carlo simulated $\gamma$-ray response functions for the $^{88}$Y decay.}
\label{YMCres}
\end{figure}

Figures~\ref{CsMCres},~\ref{CoMCres} and~\ref{YMCres} show the comparison of the experimental and the GEANT4 Monte Carlo simulated pulse height spectra for the $^{137}$Cs, $^{60}$Co and $^{88}$Y decays, respectively. The finite resolution effects quantified in Section~\ref{eneres} have been included in the Monte Carlo data. It can be observed that the agreement between the data and the simulations is excellent. The overall differences are below the 3$\%$ uncertainty in the activity of the sources. It can be however noticed that the simulations tends to underestimate of the Compton background. Such a difference is not attributed to the effect of the supports and the surrounding materials, which were carefully included, but to the unknowns in the amount, composition and densities of some materials close to the CLYC crystal. For example, the CLYC crystal is surrounded by a layer of reflective material which is not described in detail by the drawings provided by the manufacturer and was modeled as Teflon. As described in Cano-Ott et al.~\cite{Cano98}, the difficult to know density (and thickness) of the reflector surrounding an inorganic scintillator has an impact in its response function to $\gamma$-rays and can be obtained empirically from the value which reproduces best the experimental data.

\begin{figure}[h!]
\centering
\includegraphics[width=1.\columnwidth]{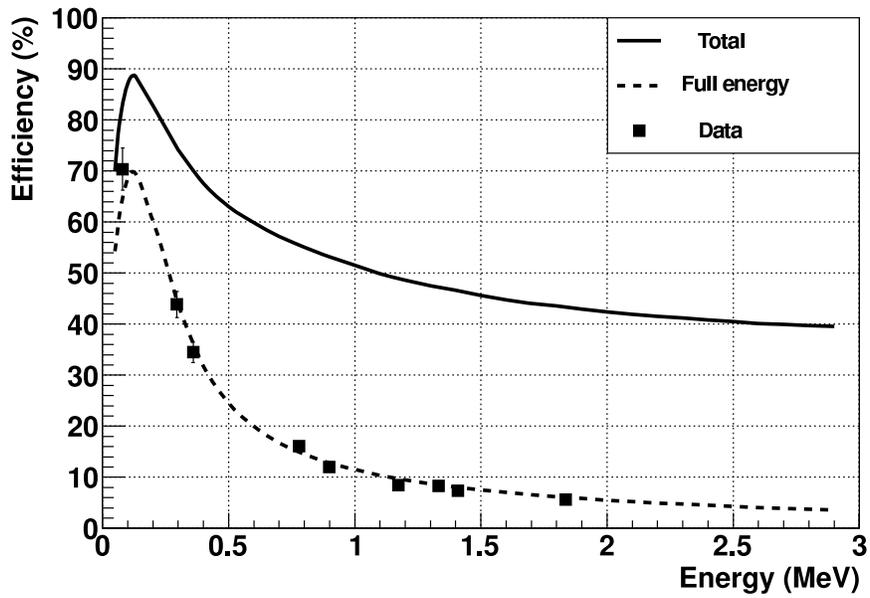}
\caption{Monte Carlo simulated intrinsic total (solid line) and full energy peak (dashed line) efficiencies compared to the experimental full energy peak data (squares).}
\label{efficiency}
\end{figure}

Fig.~\ref{efficiency} shows the intrinsic total and peak efficiencies obtained by Monte Carlo simulation in the energy range between 50 keV and 3 MeV. The agreement with the data for $\gamma$-rays (squares) coming from the decays of $^{60}$Co, $^{88}$Y, $^{133}$Ba and $^{152}$Eu improves significantly the previous work done by H.S. Kim et al.\cite{Kim}.

\subsection{n/$\gamma$ discrimination}

\begin{figure}[h!]
\centering
\includegraphics[width=1.\columnwidth]{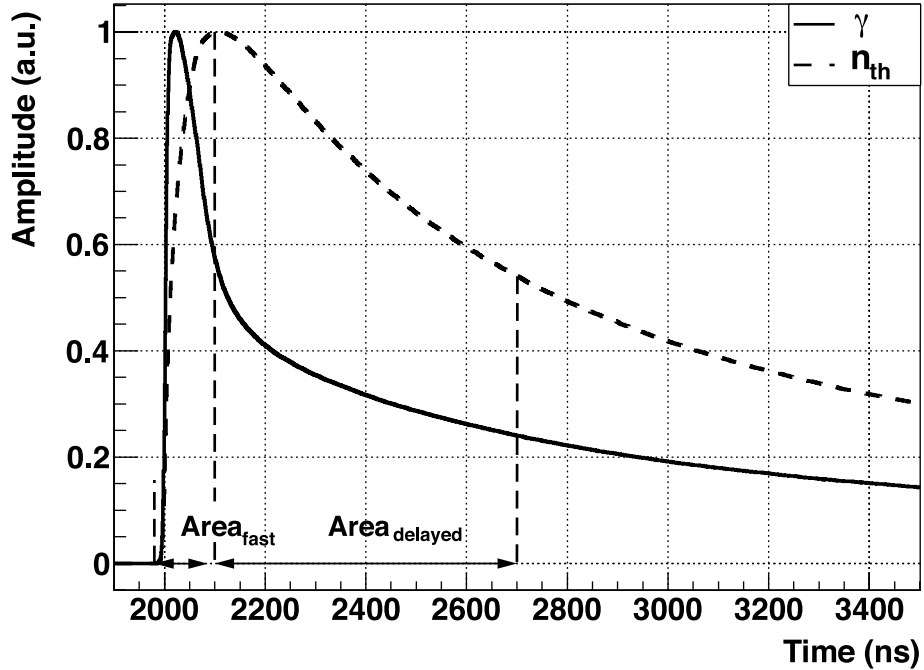}
\caption{Neutron and $\gamma$-ray averaged pulses. Blue dashed lines are used to mark the fast and delayed integration windows.}
\label{pulseshape}
\end{figure}

The discrimination between neutron and $\gamma$-ray pulse shapes has been performed using the charge integration method. The pulse integrals in two different time windows, labelled $A_{fast}$ and $A_{delayed}$, have been used to separate the two particle types. The best performance was obtained using a PSD parameter defined as $PSD={A_{delayed}}/{A_{fast}+A_{delayed}}$ \cite{Lee}, where the fast integration window starts at the beginning of the pulse and the delayed window starts at the end of the fast window. Fig.~\ref{pulseshape}. shows the average neutron and $\gamma$-ray pulse shapes, as well as the fast and delayed integration windows selected for computing the value of $PSD$. 

The performance of the discrimination has been quantified via the figure-of-merit (FOM)

\begin{equation}
FOM=\dfrac{\Delta}{\Gamma_{n}+\Gamma_{\gamma}}
\end{equation}

where $\Delta$ is the separation between the centroids of the neutron and the $\gamma$-ray PSD distributions peaks, and $\Gamma_{n}$ and $\Gamma_{\gamma}$ are the corresponding FWHMs at a given energy. The optimal value of the FOM was found for a width of the fast and delayed windows of 100 and 600 ns respectively. The energy was calculated from the integral of the pulse over 8 $\mu$s total window.

\begin{figure}[h!]
\centering
\includegraphics[width=1.\columnwidth]{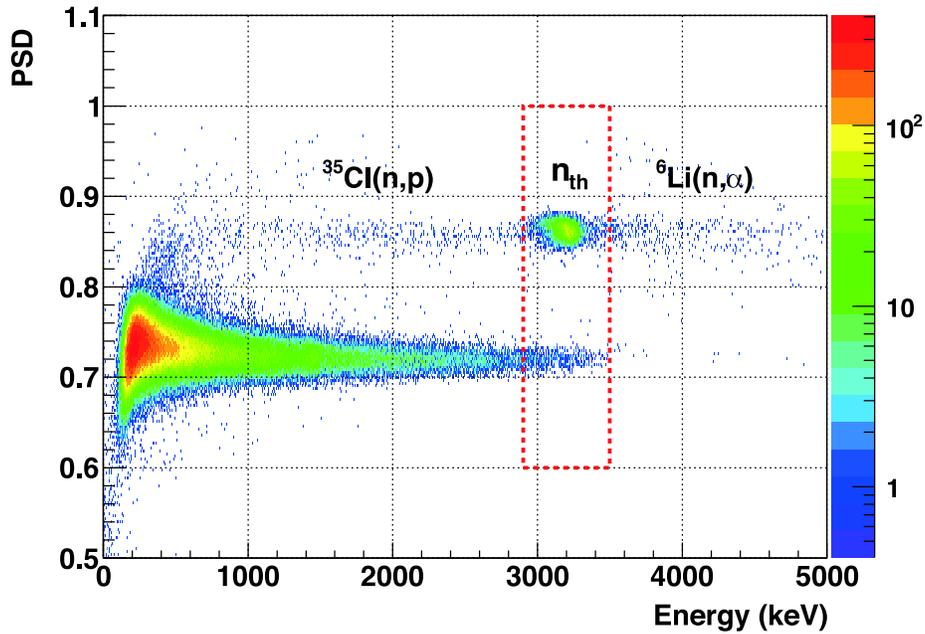}
\caption{2D PSD plot from a moderated $^{252}$Cf source. Squared region with dashed line is used to compute the FOM}
\label{ngamma}
\end{figure}

Data taken with a $^{252}$Cf source embedded in a polyethylene container were processed with the PSD algorithm. Fig.~\ref{ngamma} shows the characteristic 2D discrimination histogram where the PSD is plotted versus the energy of the signal. The $\gamma$-ray events are distributed over a wide energy range at low PSD values, while neutron events can be identified at higher PSD values. The region where the secondary particles from neutron interactions appear can be divided in three zones, the dominant structure corresponding to thermal neutrons at a $\gamma$-ray equivalent energies (GEE) around 3.2 MeV, fast neutrons from the $^{35}$Cl(n,p) reaction at lower energies and fast neutrons from $^{6}$Li(n,$\alpha$) reaction at higher energies.   

\begin{figure}[h!]
\centering
\includegraphics[width=1.\columnwidth]{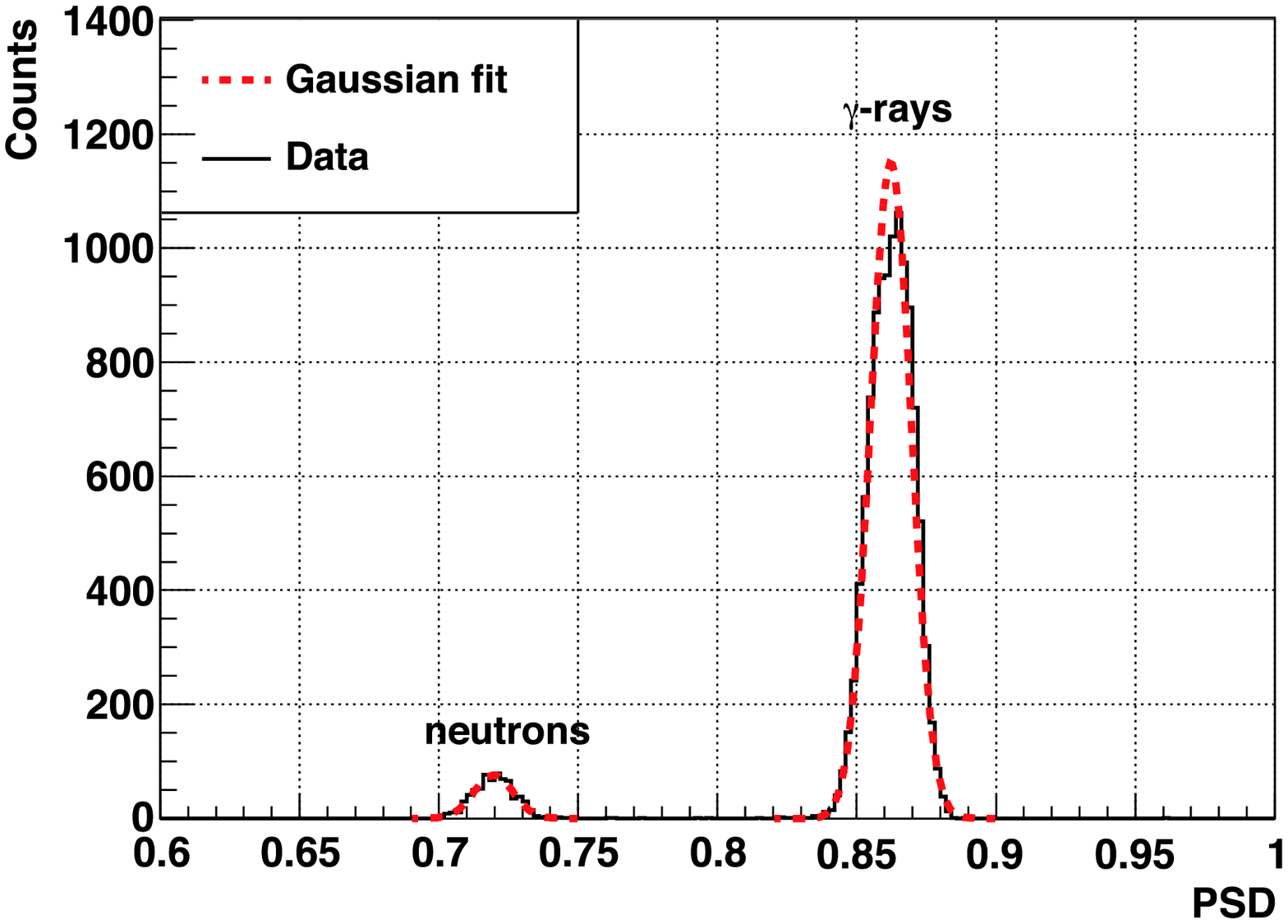}
\caption{Histogram of the PSD values after applying the event selection criteria. The data are represented in black and the red line shows the results of a Gaussian fit to the neutron and $\gamma$-ray distributions.}
\label{fom}
\end{figure}

The projection onto the PSD axis of the selected events in the marked energy region has been used to quantify the figure of merit (see Fig.~\ref{fom}). Two Gaussian functions have been fitted to the well separated peaks. A FOM of 4.2$\pm$0.1 have been obtained from fit parameters, i.e. centroids and FWHMs. The FOM value found in this work agrees with the values reported in the literature~\cite{Oly2, Lee}.  

The effect on the quality of the n/$\gamma$ discrimination of the digitalization procedure and software reconstruction has been investigated by recording various data sets with the  $^{252}$Cf source at different sampling rates of 1 GS/s and 250 MS/s and full vertical input scales. The FOM value was calculated for each configuration and a summary of the results is given in Table~\ref{tabfom}. A worsening of the FOM with decreasing sampling rate is clearly observed. It has been attributed to the degradation of the performance of the integration method with the decreasing umber of points available for calculating the baseline and the fast and delayed areas. The variation of the input scale, i.e. the decrease in the sampling resolution, did not show any sizable degradation of the FOM in the range studied.

\begin{table}[!htb]
\centering
\begin{tabular}{@{} c @{\kern.7em} c @{\kern1em} c @{\kern0.7em} c @{}} \hline 
Sampling rate & \multicolumn{3}{c}{Full scale (mV) } \\
              & 100 & 500  & 1000  \\ \hline
1 GS/s & 4.2(1) & 4.2(1) & 4.2(1) \\
250 MS/s & 3.6(1) & 3.7(1) & 3.7(1) \\ \hline
\end{tabular}
\caption{Values of the figure of merit for different setting of the waveform digitizer.}
\label{tabfom} 
\end{table}

\section{Intrinsic background}

\begin{figure}[h!]

\centering

\includegraphics[width=1.0\columnwidth]{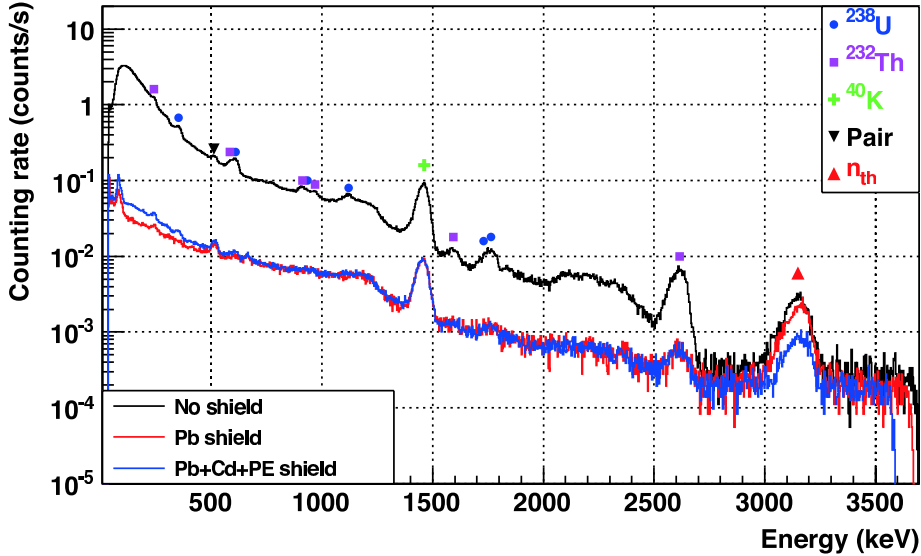}

\caption{Background spectra measured at the CIEMAT surface laboratory with different shielding configurations: without any shielding (black), with 5 cm thick Pb shielding (red) and with 5 cm Pb,1 mm Cd and 5 cm HDPE shielding (blue). }

\label{bckg}

\end{figure}

In order to have a first glimpse on the intrinsic background produced by the contamination of radioactive isotopes (mostly $^{238}$U, $^{232}$Th and $^{40}$K) in the detector materials, data have been taken with the CLYC placed inside different shielding configurations, following the procedure described in~\cite{Gia}. The spectra, recorded with the detector located inside a 5 cm thick lead castle (10 h), or inside a shield made of a 5 cm thick lead, 1 mm cadmium and 5 cm thick polyethylene (20 h), have been compared with the spectra taken with no shielding (10 h). Fig.~\ref{bckg} shows the three background spectra, normalized to the live time, recorded with the different configurations. A reduction of one order of magnitude in the background is observed for the $\gamma$-ray peaks when the Pb shield is used. Moreover, the ratio between the ($^{40}$K) 1461 keV and the ($^{208}$Tl) 2615 keV $\gamma$-ray peaks suggests its external natural origin. In addition, the neutron capture peak at 3.2 MeV is changed slightly, due to low neutron absorption cross section of the lead.   

The addition of the Cd and PE layers to the shielding does not affect significantly the $\gamma$-ray background but reduces the neutron capture peak. The enhancement in the low energy part of the spectrum is attributed to the $\gamma$-rays emitted in the neutron capture in Cd and attenuated by the lead. 

With the aim of having a more complete assessment of the intrinsic background, evidencing the contribution given by the radioisotopes in the detector materials, data have been taken in the hall A of the LSC. The radioactive contamination has been denoted by comparing the spectra taken with and without a massive 50 cm thick high density PE shield, whose contamination, measured with a germanium crystal, is a few ppb in $^{238}$U and $^{232}$Th. Data have been taken with the CIEMAT data acquisition system based on the SP ADQ14DC waveform digitizer and the recorded signals were processed with the same PSA algorithms used at the CIEMAT nuclear instrumentation laboratory. The CLYC n/$\gamma$ separation was verified in place with a $^{252}$Cf source, obtaining a figure of merit of ~4, similar to the one obtained on surface.

\begin{figure}[h!]
\centering
\includegraphics[width=1.0\columnwidth]{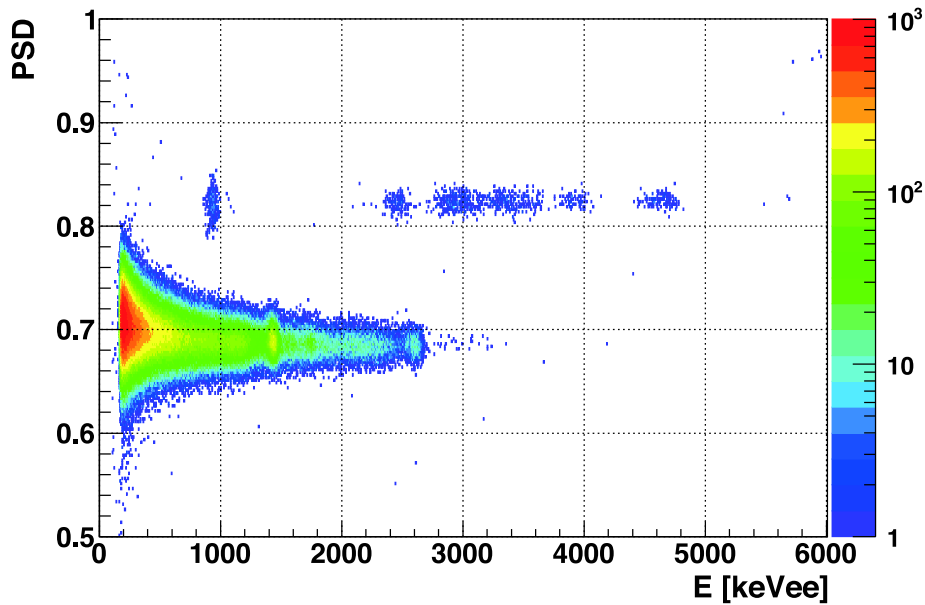}
\caption{2D plot of intrinsic background with the possible alpha peaks measured at the LSC laboratory.}
\label{intbckg}
\end{figure}

Data have been also taken with an $^{3}$He detector (model 252231 manufactured by LND Inc.~\cite{LND}), in order to measure the thermal neutron rate in the same position of the hall A and inside the shielding. The measured thermal neutron flux was obtained knowing the intrinsic efficiency of the $^{3}$He tube. The analysis of the $^{3}$He data confirmed that the number of thermal neutrons inside the PE shielding can be considered negligible for a few hours of data taking. The absence of a visible neutron peak in the pulse height spectrum when the $^{3}$He counter is placed inside the polyethylene is compatible with this hypothesis within the given statistical uncertainty.

\begin{figure}[h!]
\centering
\includegraphics[width=1.0\columnwidth]{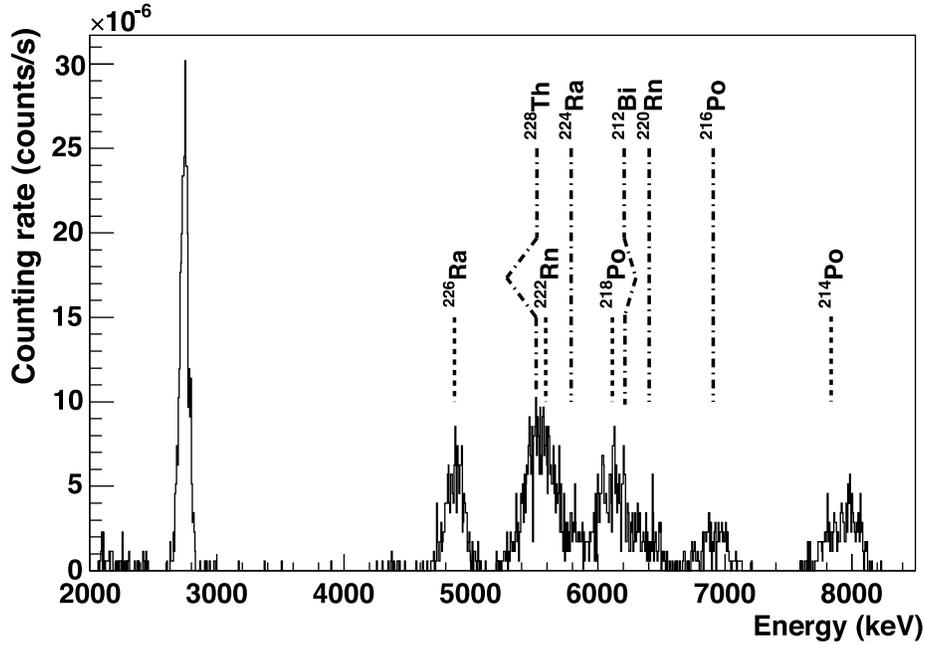}
\caption{Background spectra measured at the LSC laboratory with a tentative identification of the alpha peaks.}
\label{alph}
\end{figure}

The analysis of the data taken with the CLYC placed inside the PE shield has evidenced physics signals present in the neutron region of the 2D n/$\gamma$ discrimination plot, overlapping with the region where the neutron-induced signals are observed (Fig.~\ref{intbckg}). The data measured with the detector outside the polyethylene shield show very similar structures in the same region, in terms of energy and rate, suggesting the possible internal origin of these events. The well-resolved peaks in the pulse height spectra (see Fig.~\ref{alph}) have been interpreted as a high-energy $\alpha$-activity inside the crystal, most likely coming from the natural decay chains. Given the short half-lives of the $^{222}$Rn daughters, it is not possible to break the equilibrium of that part of the chain until the  $^{210}$Pb decay, thus, a preliminary energy calibration has been obtained identifying the $^{222}$Rn, $^{218}$Po and $^{214}$Po decays. The energy calibration has been obtained considering that the total energy deposited inside the crystal was given by the Q-value of the reaction, assuming a roughly linear response of the detector between 5 MeV - 7.6 MeV .  

Independently, it has been possible to confirm the identification of the $^{214}$Po peak through a time-resolved analysis of the correlation between the decays. Given the relatively short (164 $\mu$s) half-life  of the $^{214}$Po, the coincidence within a few hundreds of $\mu$s between a $\beta$ and $\alpha$ signals leads unambiguously to the identification of the $^{214}$Bi and $^{214}$Po decays.

The $\approx$ 10$^{-3}$ counts/s rate, calculated for both the $^{226}$Ra and $^{214}$Po decays confirms the equilibrium of that part of  the chain. The corresponding measured activity evidences a (possible bulk) contamination of ~3 mBq/kg. Such an intrinsic activity sets a lower limit to the neutron rate detectable with the CLYC and can limit its practical use in low background environments like the underground laboratories.

The interpretation of the peak at 2.7 MeV is less evident. A visual inspection of the pulses confirmed that all the events but one (a rare pile up event) had the same shape than the ones in the range between 50 and 110 area units. At present their cause is unknown but if corresponding to alpha-particles, they should be produced by alpha emitters with lower Q-values. A possibility would be a contamination of rare earths chemically compatible with the Yttrium.

\section{Conclusions}
The characterization results of a 2"x2" CLYC detector has been reported. The energy resolution has been studied in the energy range up to 2 MeV. A resolution of 4.7$\pm$0.1$\%$ has been measured for the 662 keV in $^{137}$Cs. 

Good time resolution of 1340$\pm$6 ps FWHM was obtained at $^{60}$Co energies. The present results showed that spectroscopy Super Bialkali PMT provide as good timing resolutions than fast photomultiplier tubes. Variation in the gain has been studied as a function of the irradiation point, temperature and counting rate. Rather small variation less than 0.2 $\%$ has been observed with temperature and counting rate in the range studied. On the other hand, variation at the level of 1$\%$ in gain has been observed as a function of the irradiadion point.

The response function of CLYC detector to $\gamma$-ray radiation has been calculated using the simulation GEANT4 toolkit. The simulated responses have been compared to experimental data. An excellent agreement has been obtained over the full range of energy deposited. Moreover, the experimental efficiency yielded an excellent agreement.

The PSD performance was studied with a $^{252}$Cf source. A good FOM value of 4.2$\pm$0.1 was obtained for fast and delayed signal integration windows of 100 ns and 600 ns, respectively.

A background measurement performed at the LSC evidences a ~3 mBq/kg bulk contamination. Such an intrinsic activity sets a lower limit to the neutron rate detectable with the CLYC and can limit its practical use in low background environments like the underground laboratories.

The performance of a fast waveform digitizer was investigated in terms of energy resolution, time resolution and n/$\gamma$ separation with various types of pulse shape analysis algorithms. The best energy resolution was obtained with shaping algorithm based on a CR-(RC)$^4$ filter. Moreover a large dependence on the vertical resolution (i.e. number of bits) has also been observed. A digital constant fraction discriminator algorithm lead to slightly worse results than the ones obtained with an analogue chain. Last, but not least, the quality of the n/$\gamma$ separation (PSD) as a function of the sampling rate and vertical resolution. The best PSD results were obtained at 1 Gsample/s and without any observed dependence on the vertical resolution explored. As a general conclusion, a data acquisition system based on a 14 bit and 1 Gsample/s digitizer leads to equivalent data to the ones obtained with a more voluminous and less flexible analogue electronics chain.

\section*{Acknowledgements}
This work was supported partially by the Spanish Ministry of Economy, Industry and Competitiveness - MINECO and its Plan Nacional de I+D+i de F\'{i}sica de Part\'{i}culas projects FPA2014-53290-C2-1-P,  FPA2015-70657P and FPA2016-76765-P, the "Unidad de Excelencia Mar\'{i}a de Maeztu: CIEMAT - F\'{i}sica de Part\'{i}culas" through the grant MDM-2015-0509, the European Commission CHANDA project FP7-Fission-2013-605203 and the Aragon Government.

\end{document}